# Determination of III-V/Si absolute interface energies: impact on wetting properties

S. Pallikkara Chandrasekharan[1], I. Lucci[1], D. Gupta[1], C. Cornet[1], and L. Pedesseau[1,*]

[1]*Univ Rennes, INSA Rennes, CNRS, Institut FOTON – UMR 6082, F-35000 Rennes, France*
*Corresponding author: laurent.pedesseau@insa-rennes.fr*

Here, we quantitatively determine the impact of III-V/Si interface atomic configuration on the wetting properties of the system. Based on a description at the atomic scale using density functional theory, we first show that it is possible to determine the absolute interface energies in heterogeneous materials systems. A large variety of absolute GaP surface energies and GaP/Si interface energies are then computed, confirming the large stability of charge compensated III-V/Si interfaces with an energy as low as 23 meV/Å$^2$. While stable compensated III-V/Si interfaces are expected to promote complete wetting conditions, it is found that this can be easily counterbalanced by the substrate initial passivation, which favors partial wetting conditions.

## I. INTRODUCTION

The epitaxial integration of dissimilar materials with different physical properties is nowadays one of the greatest challenges in materials science as it enables the development of novel multi-functional devices. A large variety of materials associations have been attempted by several groups, e.g. oxides/III-V,[1] oxides/Si, [2] III-V/Si, [3] III-V/Ge [4], hybrid perovskites/PbS [5]. But it is also well known that the performances of photonic, or energy harvesting devices developed with such heterogeneous materials associations strongly depend on materials defects, that are generated at the hybrid interface. [6–10] Fundamentally, at hybrid interfaces, the mechanical stacking of atoms is constrained by the chemical mismatch between the atoms of the substrate and those of the deposited layer. The real atomic arrangement at hybrid interfaces is very hard to determine experimentally, although it has drastic consequences on its physical properties, such as electronic wavefunctions spatial extents, dielectric, vibrational or transport properties and band bendings near the interface (see e.g. [11] ). Nevertheless, ab initio atomistic calculations have previously been used to assess the relative stability of several comparable heterogeneous interfaces in the III-V/Si model case: GaP/Si [12–14]. Especially, the relative stability of abrupt III-Si or V-Si interfaces and charge-compensated interdiffused interfaces (where group III or group V atoms are mixed with Si ones within one or few monolayers) was discussed in details. [11–13,15] In these previous works, the authors explained the good stability of charge-compensated interfaces by using the electron counting model (ECM) criteria [16]. A very recent analysis of the high-resolution Transmission Electron Microscopy of the GaP/Si interface confirmed the probable formation of Ga-compensated interfaces [17]

This approach is particularly interesting, as it challenges the conventional description of III-V/Si epitaxial integration which usually considers that once a monoatomic layer is sent to the substrate surface, it remains sticked without any evolution or change during subsequent layers growth. But at typical epitaxial temperatures (above 400°C), the reorganization of the interface atomic arrangement is indeed very probable.

While the authors previously pointed out a very important criterion for interface stability [8–10,12], the lack of any absolute quantitative estimation of the interface energy hampers the clarification of partial/total wetting conditions, that strongly impacts the defects generation [6]. Indeed, the determination of the Young-Dupré spreading parameter $\Omega$ [18] requires the quantitative knowledge of both surface and interface energies, which was preliminary done for abrupt interfaces only [6]. In the case of III-V/Si this parameter is given by [6]:

$$\Omega = \gamma^S_{(Si)} - \gamma^S_{(III-V)} - \gamma^i_{(III-V/Si)} \quad (1)$$

where $\gamma^S_{(III-V)}$ and $\gamma^S_{(Si)}$ are the surface energies of the most stable III-V facet that would be involved in the 2D growth on the substrate and of the silicon surface respectively and $\gamma^i_{(III-V/Si)}$ is the interface energy between the III-V semiconductor and the Si. A positive value of $\Omega$ corresponds to perfect wetting conditions, while a negative value corresponds to partial wetting, i.e., a Volmer-Weber growth, or perfect non-wetting conditions.

In this article, we first show how absolute interface energies can be computed for heterogeneous materials associations by using Density Functional Theory (DFT). We then determine the absolute interface energies for different abrupt and compensated GaP-Si interfaces. We finally discuss the impact of the interface atomic structure on the wetting properties of III-V semiconductors on Si.

## II. GENERAL STRATEGY

Here, we propose to use DFT to evaluate numerically the absolute interface energy. To this aim, two strategies can be followed: (i) hybrid materials superlattices, which reduces computational time, but also reduces the degrees of freedom for exploring different interfaces configurations, as the super-cell has to be built with two perfectly similar A/B and B/A interfaces between materials A and B. (ii) Full hybrid combination of materials with free surfaces. This solution is the most flexible one, as all the possible interfaces can be analyzed, whatever the crystal orientation, and the atomic structure of the interface. This however requires heavy DFT calculations and large computational times. In addition, it also requires a good knowledge of the free surfaces reconstructions and corresponding energies. But, since the knowledge of surface energies is required in any case for the analysis of wetting properties, we here adopt this second solution for GaP/Si. Fig. 1 gives an illustration of a typical supercell used for the calculations. The slab should include the bottom material (that mimics the substrate), with a free surface at the bottom, the top material (that represents the materials deposited on the substrate), with a free surface at the top, and a sufficient thickness of vacuum to avoid any electrostatic charge-charge and dipole-dipole interactions between the slab and its images. The interface is the region between the two different materials. Here, we point out that the implementation of such a method does not require to specify if one given atom belongs to the bulk, to the surface or to the interface. The only constraint is to fix the positions of some atoms in the middle of the bulk in order to avoid strain fields that could be induced by the presence of the surrounding free surfaces and the interface.

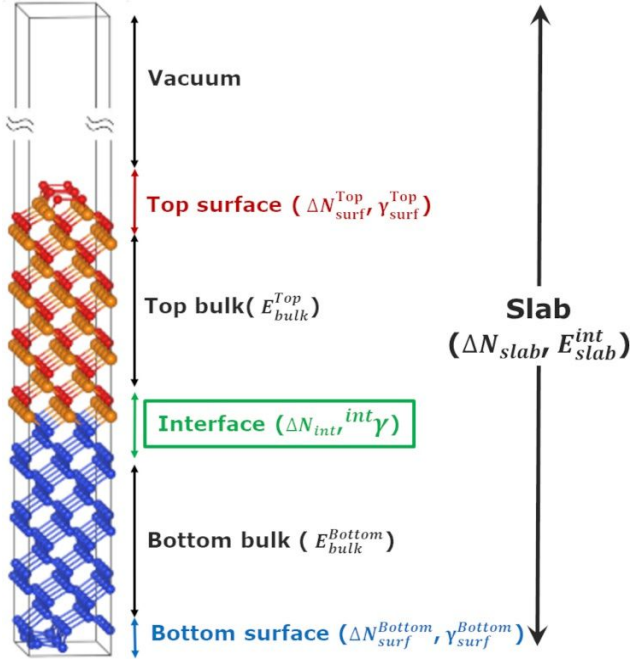

FIG. 1: Schematic of a supercell used for absolute interface energy determination, where the most important energy contributions $E$ and $\gamma$ are identified. Excess or lack of atoms induced by the surfaces and the interface as compared to the bulk, is given by the deviation to the bulk stoichiometry $\Delta N$.

We define the absolute interface energy $^X\gamma_Z^Y$, where $X$ specifies the studied interface. $Y$ and $Z$ are the top and bottom specific surfaces of the slab related to the two considered materials. The absolute interface energy is thus given by:

$$^X\gamma_Z^Y = \frac{E_{slab}^{int} - \sum_i(N_i\mu_i^{i-bulk}) - \sum_{j=Y,Z}(A\gamma_{surf}^j)}{A} \quad (2)$$

where $E_{slab}^{int}$ is the total energy of the slab, $\mu_i^{i-bulk}$ are the chemical potentials of the $N_i$ atoms composing the material $i$, $\gamma_{surf}^j$ are the surface energies of surfaces $Y$ and $Z$ (the top and bottom specific surfaces), and $A$ is the in-plane surface area of the slab. Interestingly, Eq. (2) gives an explicit dependency of the interface energy with the chemical potential and the specific surface energies that have been chosen. In our approach which is shown in Fig.1, the slab, surfaces, and interface have their own stoichiometries, i.e. their number of charges differ from the bulk by a quantity $\Delta N$ (difference between the number of P atoms and Ga atoms) referred hereafter as surface or interface stoichiometries. So, at the end, the overall deviation to the bulk stoichiometry $\Delta N$ for the slab is dependent of the respective stoichiometries of the interface and of the two surfaces. Only a careful analysis of the individual stoichiometries allow to determine absolute surface and interface energies independently. More details about stoichiometry for surface and interface are given in Supplementary Materials section 2. [19] This will be shown hereafter. Furthermore, a specific surface or interface stoichiometry reduced parameter $S$ is defined (Eq. (S1, S2)) as the number of P atoms minus the number of Ga atoms ($\Delta N$) per (lxl) unit cell ($a_o^2$). This parameter is defined for surface and interface in Supplemental Materials, section 2. [19]

In the following, we will apply this procedure to the quasi-lattice-matched GaP/Si model case. While the silicon is a homovalent crystal with a diamond structure, GaP and many III-V semiconductors are heterovalent crystals with a Zinc-Blende structure, that leads to many possible interface atomic configurations.

### III. THE GaP/Si ABSOLUTE INTERFACE ENERGY

The calculations were performed within the DFT [20,21] as implemented in SIESTA package [22,23] with a basis set of finite-range of numerical atomic orbitals. Calculations have been carried out with the generalized gradient approximation (GGA) functional in the Perdue-Burke-Ernzerhof (PBE) form [24] Troullier–Martins pseudopotentials [25], and a basis set of finite-range numerical pseudo-atomic orbitals for the valence wave functions [26].

### A. Si and GaP surfaces

Although the precise determination of Si and GaP surface energies is first motivated by the absolute interface energy calculation, we provide here a detailed analysis of surface energies for a complete set of stable facets, including the (001) Si and GaP surfaces and higher index (114), (2 5 11) and (111) facets, that have been already observed experimentally for GaP [6,27]. The charge balances for these surfaces were carefully analyzed, and were found to fulfill the electron counting model (ECM) stability criteria. [16]

The general equation used to determine the surface energy is:

$$\gamma_{surface} A = E_{slab} - \sum_i \mu_i N_i \quad (3)$$

where $\gamma_{surface}$ is the surface energy, $A$ the surface area, $E_{slab}$ is the slab energy calculated by DFT (after relaxation calculations), $\mu_i$ is the chemical potential of the species i and $N_i$ is the number of particles of the species i in the slab. The temperature dependence is ignored because the contributions tend to cancel for free-energy differences as claimed for the GaAs material in a previous study [28]. This relation will be upgraded depending on the different surface polarity.

Extensive works were performed by DFT to study the stability of the different Si surfaces. Among all the stable reconstructions of the flat Si(001) surface, the minimum minimorum of surface energy is the c(4x2), where c stands for centered because the dimers are centered buckled in a (4x2) configuration [29,30]. To study the surface energy associated to this reconstruction, we built a periodic slab on the [1-10] and the [110] directions, whose top surface is orthogonal to the [100] direction. The slab is composed by a Si bulk sandwiched between two symmetric surfaces (see Fig 2 (b)). In the cross-sectional view presented in Fig. 2 (a), it is possible to have a better view of the buckling behavior of the dimers, while in Fig. 2 (c) the c(4x2) primitive cell is highlighted by dashed lines reconstruction top view.

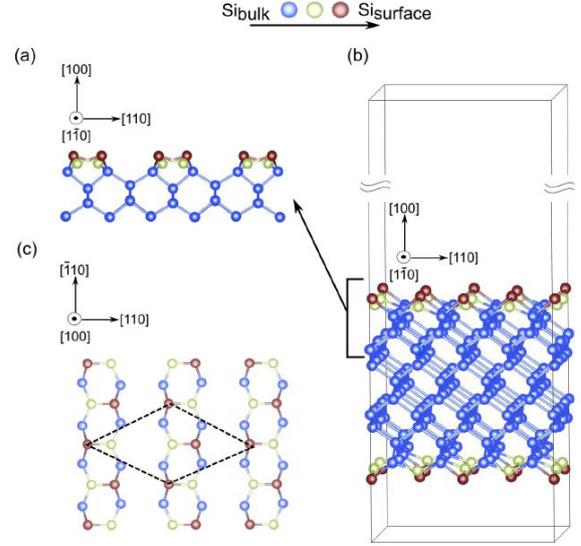

FIG 2: The Si(001) surface studied by DFT with (a) side view of the c(4x2) reconstructed surface (b) slab with symmetric surfaces realized for DFT calculations (c) surface top view with the primitive cell indicated by dashed lines.

From the general equation Eq. (3), considering that the top and bottom surfaces have the same surface energies (because it is a non-polar surface), the silicon surface energy is given by Eq. (4):

$$\gamma_{Si} = \frac{E_{slab} - N_{Si} \mu_{Si-bulk}}{2A} \quad (4)$$

where $N_{Si}$ is the total number of Si atoms in the slab. The surface energy of the Si(001) c(4x2) reconstruction was found to be 92.8 meV/Å$^2$ and reported in Table I.

For GaP, the different surface and interface energies depend on the chemical potential variations. The chemical potentials $\mu_P$ and $\mu_{Ga}$ are defined as the variables that each element can have within the bulk or surface of the GaP material. The thermodynamic conditions, to which the chemical potentials have to obey to, are the following: the upper limit of $\mu_P$ and $\mu_{Ga}$ is reached when each element is in its own pure bulk phase:

$$\mu_P < \mu_P^{P-bulk} \quad (5)$$
$$\mu_{Ga} < \mu_{Ga}^{Ga-bulk} \quad (6)$$

Moreover, at thermodynamic equilibrium the sum of $\mu_P$ and $\mu_{Ga}$ must be equal to the chemical potential $\mu_{GaP}^{GaP-bulk}$ of the GaP bulk phase:

$$\mu_{Ga} + \mu_P = \mu_{GaP}^{GaP-bulk} \quad (7)$$

$$\mu_{GaP}^{GaP-bulk} = \mu_{Ga}^{Ga-bulk} + \mu_P^{P-bulk} + \Delta H_f(GaP) \quad (8)$$

where $\Delta H_f(GaP)$ is the heat of formation of GaP material. $\mu_P^{P-bulk}$ and $\mu_{Ga}^{Ga-bulk}$ are the chemical potentials of the species P and Ga at which the black P and α-Ga phase can form respectively. In this work, the value of -0.928 eV for $\Delta H_f(GaP)$ has been determined from the calculated values above, in agreement with the literature [31–33].

Here, we therefore calculate the GaP(001), GaP(114), GaP(2 5 11), and GaP(111) surface energies and the GaP/Si interface energy as a function of the phosphorus chemical potential variation $\Delta\mu_P = \mu_P - \mu_P^{P-bulk}$. Thus, by combining (7), (8), (9) and (10), the extreme thermodynamic conditions for $\Delta\mu_P$ are given by:

$$\Delta H_f(GaP) < \Delta\mu_P < 0 \quad (9)$$

Therefore, when $\Delta\mu_P$ equals the heat of formation $\Delta H_f(GaP)$, the extreme Ga-rich limit is reached (i.e., bulk Ga will form preferentially). Contrary to that, when $\Delta\mu_P$ equals 0, the extreme P-rich limit is reached (i.e., bulk P will preferentially form).

For the non-polar GaP(001) surfaces, the bottom and top surface have been treated identically with the same reconstruction which decreases the error on the determination of the surface energy. In particular, for the Ga-rich GaP(001) surface, the GaP(001)md(2x4) reconstruction [34,35] (where md stands for mixed dimers) is assumed. This reconstruction is often considered for Ga-rich conditions in the literature. [36–38] For the P–rich GaP(001)(2x4) surface, different stable structures were proposed [39]. In this work, we studied a simple anion P-rich GaP(001) (2x4) surface that fulfills the ECM criteria as proposed for GaAs [40]. The thicknesses of the slab are about 17Å and 23Å respectively for the P-rich GaP(001)(2x4) surface (Fig. 3 a,c) and for the Ga-rich GaP(001)md(2x4) surface (Fig. 3 d,f). The outermost atoms (within about 6Å from the vacuum region) of the top and bottom surfaces were allowed to relax to their minimum energy configuration and all other atoms were kept frozen in the bulk position.

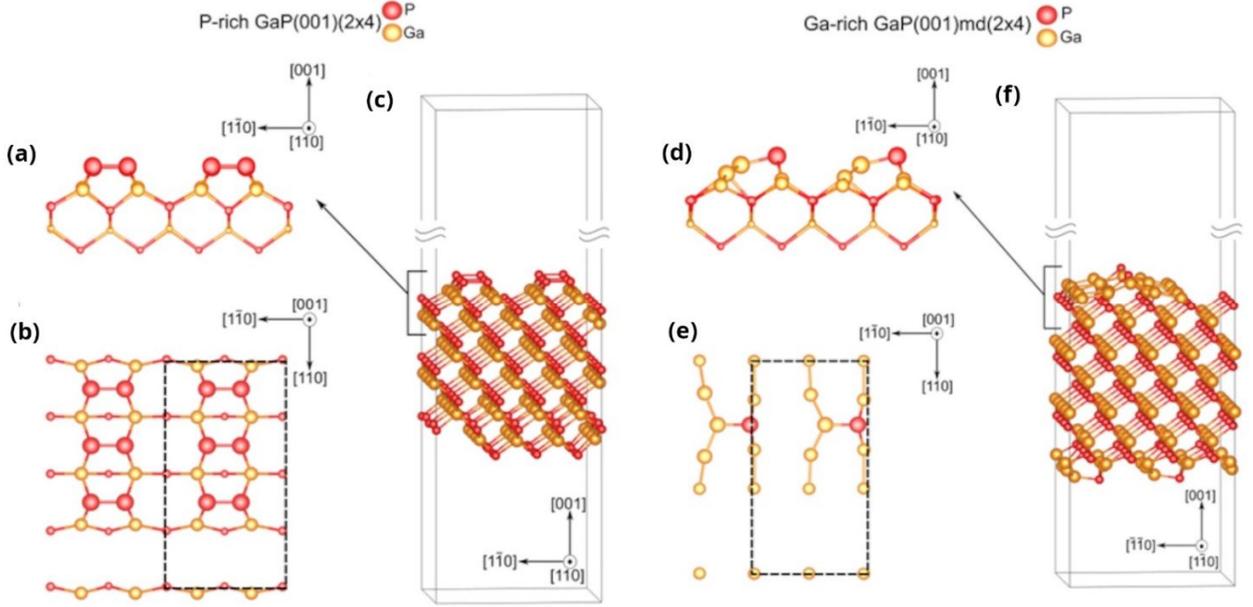

FIG. 3: (a) Side profile, (b) top view, and (c) the slab realized for P-rich GaP(001)(2x4) surface. (d) Side profile, (e) top view and (f) the slab realized for Ga-rich GaP(001)(2x4)md surface. Dashed lines in the top views indicate the unit cells of the reconstructions.

Table I. Stable GaP and Si surface energies computed by DFT and corresponding surface stoichiometries. The surface stoichiometry parameters $S_{surf}$ per (lxl) unit cell of GaP ($a_{GaP}^2$). The in-plane surface area $A$ of these slabs per (1x1) unit cell of GaP.

| Surface | $A/a_{GaP}^2$ | $\Delta N_{surf}$ | $S_{surf}$ | Energy (meV/Å²) P-rich | Energy (meV/Å²) Ga-rich |
|---|---|---|---|---|---|
| Si(001) c(2×4) | | | | 92.8 | |
| GaP(001) (2x4) | 8 | 4 | $\frac{1}{2}$ | 57.4 | 72.4 |
| GaP(001) md (2x4) | 8 | -8 | -1 | 82.8 | 52.9 |
| GaP(114)B α2(2x1) | $3\sqrt{2}$ | 0 | 0 | 67.3 | 67.3 |
| GaP(114)A α2(2x1) | $3\sqrt{2}$ | 0 | 0 | 59.9 | 59.9 |
| GaP(2 5 11)B (1x1) Ga-dimer | $\frac{5\sqrt{7}}{2}\cos(22.21)$ | +2 | $\frac{1}{\sqrt{150}}$ | 66.8 | 58.3 |
| GaP(2 5 11)A (1x1) P-dimer | $\frac{5\sqrt{7}}{2}\cos(22.21)$ | -2 | $-\frac{1}{\sqrt{150}}$ | 52.9 | 62.7 |
| GaP(111)A Ideal | $2\sqrt{3}$ | -2 | $-\frac{1}{\sqrt{3}}$ | 95.7 | 78.4 |
| GaP(111)B Ideal | $2\sqrt{3}$ | +2 | $\frac{1}{\sqrt{3}}$ | 75.1 | 92.3 |
| GaP(111)A Ga-vacancy | $2\sqrt{3}$ | 0 | 0 | 51.2 | 51.2 |
| GaP(111)B Ga-adatom | $2\sqrt{3}$ | 0 | 0 | 57.3 | 57.3 |
| GaP(111)A Ga-trimer | $2\sqrt{3}$ | -8 | $-\frac{4}{\sqrt{3}}$ | 106.4 | 37.3 |
| GaP(111)B P-trimer | $2\sqrt{3}$ | +8 | $\frac{4}{\sqrt{3}}$ | 20.2 | 89.3 |

The strategy used for the silicon surface is still pertinent for non-polar GaP(001), as top and bottom surfaces are the same. For this non-polar surface, again, the energy of one surface is half the total energy of the back and top surfaces. However, in this case, both the chemical potential and the stoichiometry have an impact on the surface energy variations: The surface energy is thus expressed as:

$$\gamma_{GaP(001)} = \frac{E_{slab} - N_{Ga}\mu_{GaP}^{GaP-Bulk} - (N_P - N_{Ga})\mu_P}{2A} \quad (10)$$

where $N_{Ga}$ and $N_P$ are respectively the number of Ga and P atoms in the slab. The deviation of the surface stoichiometry as compared to the one of the bulk $\Delta N_{surface}$ is reported in Table I. Surface energies are found equal to 72.4 (52.9) meV/Å² in Ga-rich conditions and 57.4(82.8) meV/Å² in P-rich conditions for the GaP(001)(2x4) surface (for the GaP(001)md(2x4) surface).

For polar (1 1 4), (2 5 11) and (1 1 1) surfaces, the previous simple approach cannot be adopted because top and bottom surfaces are different. Type A and B polar surfaces [41] were computed for each investigated surface. In particular, letter A(B) refers to P(Ga)-terminated surfaces. As a consequence, a different approach has been applied for these cases. It consists of considering the top surface as the one being investigated and passivating the bottom one with fictitious hydrogen atoms (see Supplemental Materials-section III). [19] The method and its validity are discussed in the Supplemental Materials section IV. [19]

So, surface energies are found equal to 67.3 meV/Å² for the P-rich GaP(114)B α2(2x1) (Fig S4 (a,b,c)), and 59.9 meV/Å² for the Ga-rich GaP(114)A α2(2x1) (Fig S4 (d,e,f)). With such stoichiometric surfaces, $N_P = N_{Ga}$, and therefore the values found apply for all the range of chemical potential. These results are reported in Table I.

With the same procedure, other stable facets can be considered. Different works (e.g. for GaAs or GaP-based materials) report on the observation of III-V crystal facets lying around the (136) orientation in the stereographic triangle. [6,42–46] The needs to fulfill the ECM stability criteria led to the identification of the (2 5 11) plane as the most stable facets at the vicinity of the (136) orientation [42–44]. The two reconstructions are named respectively P-dimer GaP(2 5 11)A-(1x1) as shown in Fig. S5 (a,b,c) and Ga-dimer GaP(2 5 11)B-(1x1) as shown in Fig. S5 (d,e,f) (we use the notation of ref. [42,43]). The polar surfaces reconstructions were passivated by the fictitious H* with fractionally charged hydrogen 1.25e and 0.75e for Ga and P dangling bonds. Then, the subsurface opposite to the passivated surface of the slab was allowed to relax about 6Å into their minimum energy and all the other atoms were kept frozen in the bulk position except the fictitious H* atoms which were also allowed to relax. Similar to the (114) facet case, Eq. (S5) is used to calculate the surface energy of the two different A and B (2 5 11) facets. Surface energies are found equal to 62.7 (58.3) meV/Å² in Ga-rich conditions and 52.9 (66.8) meV/Å² in P-rich conditions for the P-dimer GaP(2 5 11)A (1x1) surface (Ga-dimer GaP(2 5 11)B (1x1)). Unlike the (114) surfaces, these reconstructions are not stoichiometric, and therefore the surface energies vary with the phosphorus chemical potential. Results are summarized in Table I.

Finally, additional calculations were performed on various GaP(111) surfaces. (111) facets are commonly identified in several works [28,41,47–59]. We first considered the two ideal surfaces namely GaP(111)A Ideal and GaP(111)B Ideal. These surface energies are found equal to 78.4 (92.3) meV/Å² in Ga-rich conditions and 95.7 (75.1) meV/Å² in P-rich conditions for the GaP(111)A Ideal surface (GaP(111)B Ideal). Then, we studied the two surfaces which are respecting the ECM by adding or removing Ga atom namely GaP(111)A Ga-vacancy as shown in Fig. S6 (a,b,c) and GaP(111)B Ga-adatom as shown in Fig. S6 (d,e,f). Here $N_P = N_{Ga}$, therefore the values of the surface energy are not evolving with the chemical potential. Surface energies are found equal to 51.2 meV/Å² for the P-rich GaP(111)A Ga-vacancy, and 57.3 meV/Å² for the Ga-rich GaP(111)B Ga-adatom. In addition, the two extreme cases where trimers of Ga or P are forming on the surface, namely GaP(111)A Ga-trimer as shown in Fig. S7 (a,b,c) and GaP(111)B P-trimer as shown in Fig. S7 (d,e,f) are also considered in this work. These surface energies are found equal to 37.3 (89.3) meV/Å² in Ga-rich conditions and 106.4 (20.2) meV/Å² in P-rich conditions for the GaP(111)A Ga-trimer surface (GaP(111)B P-trimer).

Overall, all the previously calculated surface energies are summarized in Table I. Experimentally, the growth is imposed on the specific direction (001). The appearance of other facets may result from the competition between surface energy of the (001) surface and surface energies of other stable surfaces. In fact, the possible destabilization of the (001) surface by other surfaces on a (001) substrate can be easily calculated after the pioneering works of J. Poynting and J. Thompson and others' studies [27,60,61]. Consequently, one should consider the corrected surface energy:

$$\gamma_{U/001} = \frac{\gamma_U}{\cos(\theta)} \quad (11)$$

where $\gamma_{U/001}$ is the corrected surface energy of the considered stable facet $U$, growing on a (001) substrate. In our case, $U$ is the studied surface (114), (2 5 11), or (111), and $\theta$ is the angle between $U$ and the (001) surfaces. Here, the related angles are 19.47°, 26.09° and 54.74° for (114), (2 5 11), and (111) surfaces respectively. So, the surface energies (114), (2 5 11), and (111) are divided by (Eq. 11) 0.94, 0.90, 0.58 respectively. Figure 4 shows the corrected surface energies as a function of the variation of the chemical potential P. From Ga-rich to P-rich, the most stable surfaces are GaP(001)md (2x4) for the extreme case of Ga-rich,

GaP(114)A α2(2x1) in the middle and the GaP(001) (2x4) in competition with GaP(2 5 11)A (1x1) for the extreme case of P-rich. In addition, the GaP(111)B P-trimer is even more stable than the GaP(001) (2x4) and GaP(2 5 11)A (1x1) but only in the P-rich limit.

The reconstruction of studied surfaces shows a diversity of arrangement of dimers as displayed in Fig. 3 and Fig S4-S7 and dimer lengths are also reported in Table II. The mixed dimers of Ga-P, as in Figure 3 (d, e, f), for the Ga-rich GaP(001)md (2x4) surface, are found to have a characteristic length of 2.41Å which is also the same value than the typical distance between Ga and P atoms in the bulk material. The Phosphorus dimers (in Figure 3(a,b,c), S4, S5(a,b,c) and S7(d,e,f)) have similar lengths for P-rich GaP(001) (2x4), GaP(114)A and B, GaP(2 5 11)A (1x1), and GaP(111)B P-trimer surfaces. Actually, these phosphorus dimer lengths for all of these surface reconstructions are nearly the one of the black phosphorus material. Instead, the gallium dimer lengths (Figure S4, Figure S5(d, e, f), and Figure S7(a, b, c)) are much more different and we can separate them in two categories: the shorter bond lengths for the GaP(114)B α2(2x1) and GaP(2 5 11)A (1x1) cases and the larger bond lengths for the GaP(114)A α2(2x1) and GaP(111)A Ga-trimer cases. The variation between the two categories is about 0.2Å. In fact, the Gallium dimer lengths of the Ga-rich GaP(114)A-α2(2x1) and GaP(111)A Ga-trimer surfaces are tending to be similar to the α-Ga phase. Finally, this simple analysis tends to reveal that the elastic energy should decrease for the polar surface (114) A because Ga-Ga and P-P dimer lengths are quite comparable to their length within bulk materials such as α-Ga phase and black Phosphorus. Similarly, for the (2 5 11) A and B surfaces, the surface energy decreases (increases) for the polar surface A (B) because there is less (more) difference as compared to the dimer length within bulk materials such as black phosphorus (α-Ga phase). Identically, we have the same conclusion for the polar surfaces GaP(111)A Ga-trimer and GaP(111)B P-trimer where the GaP(111)B P-trimer is more stable, also due to elastic energy which decreases as the dimers bond length are closer to the bulk black phosphorus.

Table II. Calculated dimer lengths, in units of Å, of the surface atoms for the P-rich GaP(001)(2x4), the Ga-rich GaP(001)md(2x4), the GaP(114)B-α2(2x1), the GaP(114)A-α2(2x1), the GaP(2 5 11)B (1x1), the GaP(2 5 11)A (1x1), and GaP(111)A and B reconstructions. Dimer lengths of black phosphorus, α-Ga phase, and GaP are also reported.

| Surface reconstruction and Bulk materials | Dimer length | |
|---|---|---|
| | P-P | Ga-Ga |
| P-rich GaP(001) (2x4) | 2.30±0.01 | |
| P-rich GaP(114)B-α2(2x1) | 2.32±0.02 | 2.58±0.06 |
| Ga-rich GaP(114)A-α2(2x1) | 2.30±0.01 | 2.77±0.01 |
| GaP(2 5 11)B (1x1) Ga-dimer | | 2.54±0.07 |
| GaP(2 5 11)A (1x1) P-dimer | 2.30±0.01 | |
| GaP(111)A Ga-Trimer | | 2.74±0.20 |
| GaP(111)B P-Trimer | 2.32±0.01 | |
| Black Phosphorus | 2.28±0.01 | |
| α-Ga phase | | 2.83±0.07 |
| | md Ga-P | |
| Ga-rich GaP(001) md(2x4) | 2.41±0.02 | |
| GaP | 2.41±0.01 | |

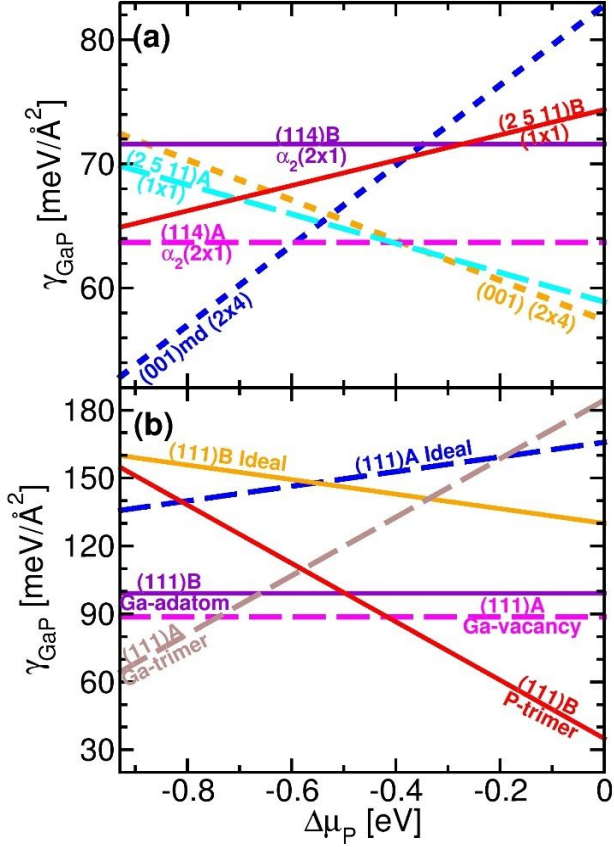

FIG. 4: Absolute surface energies of various GaP stable surfaces as a function of the chemical potential variations for (a) (001), (114), (2 5 11) GaP surfaces and b) (111) GaP surfaces. All the surface energies have been corrected from their angle $\theta$ to compare their stability on a (001) substrate.

**B. GaP/Si interfaces**

The thermodynamic analysis of the GaSiP phases is indicating a broad stability region and indicated in blue shaded ( See Supplemental materials [19] Fig S8 and S9) for the GaP/Si interface. Indeed, only two secondary phases SiP and SiP$_2$ for extreme Ga-poor and Si-rich growth conditions are in competition with the GaP/Si interface. Otherwise, GaP/Si interfaces are expected to be stable, with no secondary phase. So far, the GaP-Si hetero-interface energy has already been investigated by previous works. Indeed, results on the relative interface formation energy of the GaP on different Si surfaces has been already presented in reference [31]. The stability of the compensated GaP/Si(001) interface with respect to an abrupt one has been reported as well in references [32,62] by calculating its relative formation energy. The GaP/Si(001) absolute abrupt interface energy has also been determined [63] but these results have been questioned in ref. [64], since the dependence from the chemical potential of the absolute interface energy has not been considered. So finally, a correct determination of the different GaP/Si(001) absolute interface energies have not been proposed yet.

Our DFT calculations to determine the GaP/Si(001) absolute interface energies as a function of the chemical potential are presented in the following. The calculations computed by DFT have been done using the same parameters already reported in the paragraph above for the GaP surfaces energies study.

To determine the interface energies, we studied first the abrupt P/Si and Ga/Si interfaces (Fig. 5 (a,b)) and then the non-abrupt [62] also called compensated 0.5Si:0.5Ga-P and 0.5Si:0.5P-Ga interfaces (Fig. 5 (c,d)). For the abrupt interfaces, the slabs are shown in Fig. 6 a-d. In fact, for each interface, two different top surfaces are modeled in order to check the validity of the results and confirm that interface energies and their variations with chemical potential are independent of the top surface. Therefore, the two reconstructed GaP(001) surfaces were used for this purpose: the Ga-rich GaP(001)md (2x4) and the P-rich GaP(001)(2x4). The unit cell of each one is shown in the top view section in Fig.3. Note that, instead of keeping reconstructed GaP top surfaces and Si bottom surfaces, for the absolute interfacial energy determination, one could prefer using GaP and Si surfaces passivated with fictitious hydrogen atoms, to avoid parasitic surface/surface or surface/interface interactions. This approach is meaningful, but needs a precise evaluation of the contributions of passivated surfaces to the total energy of the slab. In this study, the large vacuum thickness, the large materials thickness and the systematic comparison of the interface energies for two very different GaP surfaces guarantee the weak contribution of these parasitic interactions. The stoichiometry of the interface for a slab is provided by the relation below:

$$\Delta N_{int} = \Delta N_{slab} - \Delta N_{surface}^{Top} - \Delta N_{surface}^{Bottom} \qquad (12)$$

where $\Delta N_{slab}$, $\Delta N_{surface}^{Top}$, and $\Delta N_{surface}^{Bottom}$ are the stoichiometry of the entire slab, the GaP surface at the top and the silicon surface at the bottom respectively. Here, the stoichiometry of the silicon surface at the bottom is equal to 0, $\Delta N_{surface}^{Bottom} = 0$, which reduces the Eq. (12). For the abrupt interfaces, the ECM is not respected (stoichiometry of +4 or -4) as shown in Table III, whereas for the compensated interfaces the ECM is respected (stoichiometry equals to 0).

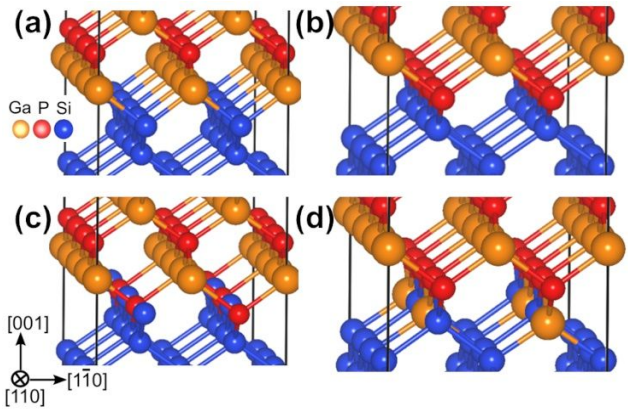

FIG 5: Zoomed atomic configuration for the two abrupt interfaces a) Si-Ga, b) Si-P, and the two compensated interfaces c) 0.5Si: 0.5P-Ga, d) 0.5Si: 0.5Ga-P, respectively.

Table III. Stoichiometry of abrupt and compensated interfaces under different surfaces, $\Delta N_{int}$. Stoichiometry of the slab, and of the top surface, $\Delta N_{slab}$ and $\Delta N_{surface}^{Top}$. The interface stoichiometry parameters per (lxl) unit cell of Si ($a_o^2$). The in-plane surface area $A$ of these slabs is $4a_o^2$.

| Slab configuration | Interface | Bottom surface | Top surface | $\Delta N_{slab}$ | $\Delta N_{surface}^{Top}$ | $\Delta N_{int}^*$ | $S_{int}$ |
|---|---|---|---|---|---|---|---|
| $^{Si-Ga}\gamma_{GaP\,(001)2x4}^{Si\,(100)}$ | Si–Ga | Si (100) | GaP (001) 2x4 | -2 | 2 | -4 | -1 |
| $^{Si-Ga}\gamma_{GaP\,(001)\,2x4md}^{Si\,(100)}$ | – | – | GaP (001) 2x4md | -8 | -4 | -4 | -1 |
| $^{Si-P}\gamma_{GaP\,(001)\,2x4}^{Si\,(100)}$ | Si–P | – | GaP (001) 2x4 | 6 | 2 | 4 | 1 |
| $^{Si-P}\gamma_{GaP\,(001)\,2x4md}^{Si\,(100)}$ | – | – | GaP (001) 2x4md | 0 | -4 | 4 | 1 |
| $^{0.5Si:0.5P-Ga}\gamma_{GaP\,(001)\,2x4}^{Si\,(100)}$ | 0.5Si:0.5P–Ga | – | GaP (001) 2x4 | 2 | 2 | 0 | 0 |
| $^{0.5Si:0.5P-Ga}\gamma_{GaP\,(001)\,2x4md}^{Si\,(100)}$ | – | – | GaP (001) 2x4md | -4 | -4 | 0 | 0 |
| $^{0.5Si:0.5Ga-P}\gamma_{GaP\,(001)\,2x4}^{Si\,(100)}$ | 0.5Si:0.5Ga–P | – | GaP (001) 2x4 | 2 | 2 | 0 | 0 |
| $^{0.5Si:0.5Ga-P}\gamma_{GaP\,(001)\,2x4md}^{Si\,(100)}$ | – | – | GaP (001) 2x4md | -4 | -4 | 0 | 0 |

Table IV. GaP/Si absolute interface energies computed by DFT.

| Slab configuration | Interface | Bottom surface | Top surface | Energy (meV/Å²) | |
|---|---|---|---|---|---|
| | | | | P-rich | Ga-rich |
| $^{Si-Ga}\gamma_{GaP\,(001)\,2x4}^{Si\,(100)}$ | Si–Ga | Si (100) | GaP (001) 2x4 | 72.0 | 40.8 |
| $^{Si-Ga}\gamma_{GaP\,(001)\,2x4md}^{Si\,(100)}$ | – | – | GaP (001) 2x4md | 69.7 | 38.5 |
| $^{Si-P}\gamma_{GaP\,(001)\,2x4}^{Si\,(100)}$ | Si–P | – | GaP (001) 2x4 | 25.9 | 57.1 |
| $^{Si-P}\gamma_{GaP\,(001)\,2x4md}^{Si\,(001)}$ | – | – | GaP (001) 2x4md | 27.3 | 58.4 |
| $^{0.5Si:0.5P-Ga}\gamma_{GaP\,(001)\,2x4}^{Si\,(100)}$ | 0.5Si:0.5P–Ga | – | GaP (001) 2x4 | 28.3 | 28.3 |
| $^{0.5Si:0.5P-Ga}\gamma_{GaP\,(001)\,2x4md}^{Si\,(100)}$ | – | – | GaP (001) 2x4md | 26.2 | 26.2 |
| $^{0.5Si:0.5Ga-P}\gamma_{GaP\,(001)\,2x4}^{Si\,(100)}$ | 0.5Si:0.5Ga–P | – | GaP (001) 2x4 | 24.7 | 24.7 |
| $^{0.5Si:0.5Ga-P}\gamma_{GaP\,(001)\,2x4md}^{Si\,(100)}$ | – | – | GaP (001) 2x4md | 22.1 | 22.1 |

Table V. Average values for stable GaP/Si absolute interface energies computed by DFT from the table IV.

| Slab configuration | Interface | Bottom surface | Top surface | Energy (meV/Å²) | |
|---|---|---|---|---|---|
| | | | | P-rich | Ga-rich |
| $^{Si-Ga}\gamma_{GaP\,(001)}^{Si\,(100)}$ | Si–Ga | Si (100) | GaP (001) | 70.9 | 39.7 |
| $^{Si-P}\gamma_{GaP\,(001)}^{Si\,(100)}$ | Si–P | – | – | 26.6 | 57.8 |
| $^{0.5Si:0.5P-Ga}\gamma_{GaP\,(001)}^{Si\,(100)}$ | 0.5Si:0.5P–Ga | – | – | 27.3 | 27.3 |
| $^{0.5Si:0.5Ga-P}\gamma_{GaP\,(001)}^{Si\,(100)}$ | 0.5Si:0.5Ga–P | – | – | 23.4 | 23.4 |

reconstructions and the P-Si interface compensated with Ga named 0.5Si:0.5Ga–P interface with g) P-rich GaP(001)(2x4) and h) Ga-rich GaP(001)md (2x4) surfaces reconstructions.

The slabs are separated by a vacuum region of 450Å thick. To avoid any surface/interface interaction, both the GaP and Si bulk are 20Å thick each. More precisely, the slabs lengths in Fig. 6 (a,e) and (b,f) are respectively 42.31Å, 43.62Å while the slabs in Fig. 6 (c,g) and (d,h) have respectively a length of 40.9Å and 45Å. For each slab, the basis vectors length is 15.44Å and 7.72Å. We choose the Si(001) as bottom surface for each case investigated. Finally, the entire GaP together with the two first layers of Silicon at the interface were allowed to relax, while the rest has been frozen.

The absolute interface energy $^X\gamma_Z^Y$ is finally calculated using Eq. (2) applied to the present specific case:

$$^X\gamma_Z^Y = \frac{E_{slab}^{int} - N_{Ga}\mu_{GaP}^{GaP-bulk} - (N_P - N_{Ga})\mu_P}{A} \\ \frac{-N_{Si}\mu_{Si}^{Si-bulk} - A\gamma_{surf}^{Si} - A\gamma_{surf}^{GaP}}{A} \quad (13)$$

where $E_{slab}^{int}$ is the total energy of the slab, $N_{Ga}$ and $N_P$ are respectively the number of Ga and P atoms of the slab investigated, $\mu_{GaP}^{GaP-bulk}$ and $\mu_P$ are the chemical potentials of the GaP bulk and of the species P. $A$ is the rectangular base surface area. $\mu_{Si}^{Si-bulk}$ is the chemical potential of the silicon bulk while $N_{Si}$ is the number of silicon atoms. $\gamma_{surf}^{Si}$ and $\gamma_{surf}^{GaP}$ are the specific bottom and top surface energies per unit area. For interfaces, the chemical potential of species P varies in the same interval range than the GaP surfaces case.

The results are reported in Table IV. The Si-Ga interface is always more stable in Ga-rich environment while the Si-P interface is more stable in the P-rich one. Moreover, as expected, this is independent of the kind of surface considered within a small numerical error. Finally, the absolute variation of the interface energy from P-rich to Ga-rich conditions is always of 31.2meV/Å² for Si-Ga and Si-P interface respectively, in agreement with the inverse stoichiometry of both interfaces. Finally, the average of stable GaP and Si absolute interface energies computed by DFT from the table IV are shown in Table V and are plotted as a function of the chemical potential in Fig. 7. The lowest value is reached (23.4 meV/Å²) with a III-V/Si Ga-compensated interface. Actually, the same evolution than the one presented in ref. [13] with relative comparisons between interfaces is observed, with an overall better stability of compensated interfaces over the whole range of chemical potential. It is also worth mentioning that the formation of the compensated interface changes drastically the absolute value of the interface energy. For instance, in Ga-rich conditions, the interface energy is divided by 2 as compared

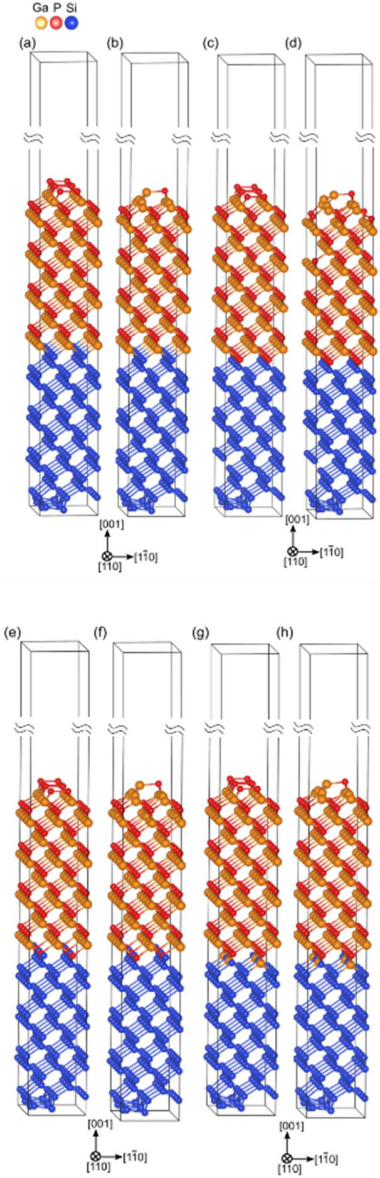

FIG 6: On the top, schematic of the four slabs used in DFT for the calculation of the absolute interface energy of the abrupt Ga-Si interface with a) P-rich GaP(001)(2x4) and b) Ga-rich GaP(001)md (2x4) surfaces reconstructions and of the abrupt P-Si interface with c) P-rich GaP(001)(2x4) and d) Ga-rich GaP(001)md (2x4) surfaces reconstructions. On the bottom, four slabs used in DFT for the calculation of the absolute interface energy of the Ga-Si interface compensated with P named 0.5Si:0.5P–Ga interface with e) P-rich GaP(001)(2x4) and f) Ga-rich GaP(001)md (2x4) surfaces

to the abrupt Ga-Si interface case. Therefore, the compensated interfaces are more likely to form during GaP/Si epitaxy and in general during III-V/Si epitaxy. We note here that the interface energies of the most stable interfaces computed in this work (between 20 and 30 meV/Å²) appear much lower than the one experimentally inferred by A. Ponchet et al. [65] that was estimated in the range of [55-110] meV/Å² for III-Sb/Si. While the change of materials system alone can hardly explain such a difference, this result raises questions about the influence of the passivation of the starting Si surface prior to the growth on the surface energy, which will be discussed later on in this work, and the contribution of interfacial defects to the interface energy, which was discussed in details in ref. [65]. Overall, the good stability of III-V/Si interfaces may change significantly the wetting properties of the system. Therefore, in the following, we examine the influence of the interface structure on the wetting properties for the GaP/Si model case.

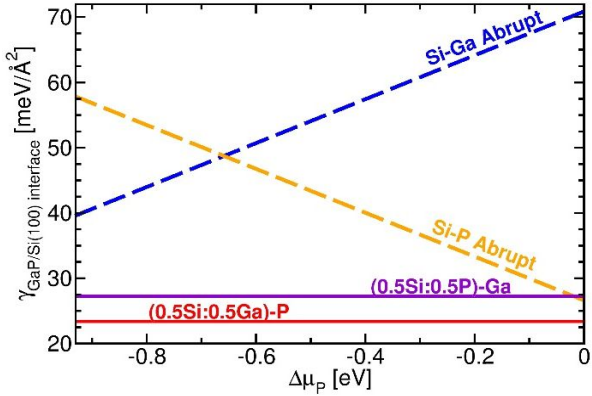

FIG 7: Average absolute interface energies diagram of both compensated and abrupt interfaces computed by DFT. As reported in [13], the compensated 0.5Si:0.5Ga-P and 0.5Si:0.5P-Ga energies are more stable with respect to the abrupt ones because they fulfill the ECM criterion. Moreover, their energy does not vary with the chemical potential $\Delta\mu_P$, as expected because of their stoichiometry ($\Delta N=0$).

## IV. GAP/SI WETTING PROPERTIES

In order to investigate the wetting properties for the GaP/Si case, the surfaces and interfaces energies computed can be used to determine the Young-Dupré spreading parameter $\Omega$ [18]. A positive value of $\Omega$ corresponds to perfect wetting conditions, while a negative value corresponds to partial wetting, i.e., a Volmer-Weber growth, or perfect non-wetting conditions, leading to the formation of 3D islands whose equilibrium shape depends on $\Omega$ [65–68]. Figure 8 represents the calculated values of $\Omega$ for the same GaP and Si surface energies, but with different interfacial atomic configurations.

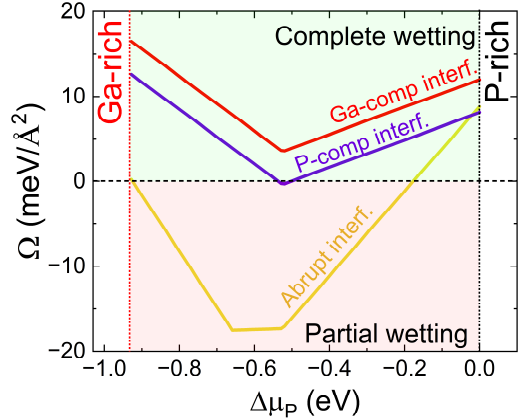

FIG 8: Young-Dupré spreading parameter $\Omega$ as a function of the chemical potential $\Delta\mu_P$, calculated for the different interface atomic configurations: (i) Abrupt III-V/Si interfaces (orange), (ii) P-compensated III-V/Si interface (purple), and (iii) Ga-compensated III-V/Si interface (red).

In this calculation, $\gamma^S_{(Si)}$ is kept constant at 92.8 meV/Å², and $\gamma^S_{(III-V)}$ is taken as the lowest GaP (001) surface energy (most stable surface) for a given chemical potential. Thus, for $\Delta\mu_P > -0.52$ eV the P-rich GaP(001)(2x4) was chosen, while for $\Delta\mu_P < -0.52$ eV the Ga-rich GaP(001)md (2x4) was alternately considered, explaining the systematic slope change at $\Delta\mu_P = -0.52$ eV observed in Fig. 8. With this approach, the spreading parameter $\Omega$ is first calculated in the case where abrupt Ga-Si or P-Si interfaces are considered (orange solid line in Fig. 8). In this case, unlike what was done in previous work [6], the most stable interface configuration for a given chemical potential was considered. Thus, for $\Delta\mu_P < -0.65$ eV, the Ga-Si abrupt interface was chosen, while for for $\Delta\mu_P > -0.65$ eV, the P-Si abrupt interface was considered. This, again, leads to a slope variation at $\Delta\mu_P = -0.65$ eV for $\Omega$. According to Fig. 8, it is clear that $\Omega$ remains negative throughout most of $\Delta\mu_P$ chemical potential range, which means that partial wetting is very likely to be achieved. The same calculation was thus performed with the most stable interface atomic configurations, corresponding to situations where the interface is either compensated by P (P-compensated III-V/Si interfaces, purple solid line in Fig. 8) or by Ga (Ga-compensated III-V/Si interfaces, red solid line in Fig. 8). In these two cases, the interface energy does not depend on the chemical potential and $\Omega$ remains positive for all the values of the chemical potential, leading to predicted complete wetting conditions. This result is theoretically not surprising, as it is a direct consequence of Eq. (1). A lower interface energy means a lower energy cost to form a 2D layer on the substrate, and thus favors the complete wetting of the material on the substrate. On the other hand, it apparently contradicts recent experimental works, which demonstrate that III-V/Si epitaxial growth happens under partial wetting conditions. [6,8] As explained previously, the wetting conditions are fully determined by the sign of $\Omega$. The extended studies conducted in this work allow to conclude that, overall, III-V (001) surface energies are lying in the [50-70] meV/Å$^2$ range, and that the III-V/Si interfaces are compensated, with energies in the [20-30] meV/Å$^2$ range. However, in the present study, a nude Si(001) surface was assumed to be representative of the substrate surface before the III-V epitaxy. As discussed in a previous work [6], the nude Si surface is very reactive because of its high surface energy, and will thus tend to passivate by creating bonds with other atoms available at the surface during chemical preparation (e.g. H, O) or annealing prior to epitaxial growth (e.g. As, P, Sb, Ga, In, Al, N). In the following, we investigate quantitatively the contribution of the substrate surface energy to the wetting properties.

Figure 9 represents the calculated values of $\Omega$ for the most stable GaP surfaces and GaP-Si interfaces (i.e., with a Ga-compensated III-V/Si interface), for different values of the substrate surface energy $\gamma^S_{sub}$, ranging from 10 meV/Å$^2$ (red solid line) to 90 meV/Å$^2$ (green solid line). It can be seen that $\Omega$ is always positive for substrate surface energies larger than 90 meV/Å$^2$. On the other hand, $\Omega$ rapidly evolves toward negative values, even for small variations of the substrate surface energy, rapidly leading to partial wetting conditions. For a substrate surface energy of 76.3 meV/Å$^2$, $\Omega$ is already negative for the whole range of chemical potential. This means that any energy gain through stabilization of the starting Si surface by more than 16.5 meV/Å$^2$ is enough to promote partial wetting conditions (i.e., growth of 3D islands). Even with an ultra-pure Si surface cleaning procedure and assuming that the epitaxial chamber is free of unintentional contaminants that could incorporate at the surface, the simple exposure of the Si surface to a group-V atoms flux at the beginning of the growth may easily stabilize the Si surface at lower values leading to partial wetting conditions. While this analysis focused on the GaP/Si case, ranges for surfaces and interfaces energies are expected to remain the same for other III-V/Si materials systems. For this reason, even if similar studies could be conducted for other III-V materials, and considering the specific surface energies of various passivated Si surfaces, the complete wetting of a Si substrate by a 2D III-V semiconductor layer appears unlikely. The detailed analysis of the Si surface passivation is consequently of great importance for the complete understanding of the system.

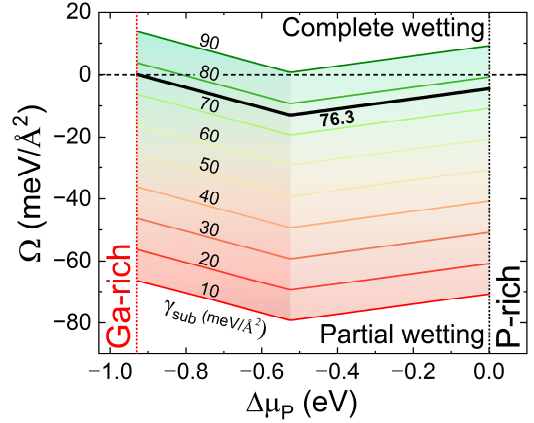

FIG 9: Young-Dupré spreading parameter $\Omega$ as a function of the chemical potential $\Delta\mu_P$, calculated for different substrate surface energies (from 10 meV/Å$^2$ (red) to 90 meV/Å$^2$ (green)), with the most stable Ga-compensated III-V/Si interface.

## V. CONCLUSIONS

In conclusion, absolute surfaces and interfaces energies were calculated for various atomic configurations of GaP surfaces and GaP/Si (001) interfaces by using density functional theory. These studies not only confirm the large stability of compensated III-V/Si interfaces but allow a quantitative analysis of the different surfaces and interfaces contributions to the wetting properties of the materials system. The large stabilization of the III-V/Si interface through charge compensation is expected to promote the complete wetting conditions in the system. We show that this effect is however easily counterbalanced by the pre-growth passivation (intentional or not) of the Si substrate surface. This quantitative analysis confirms that the complete analysis of wetting properties of III-V semiconductors on Si surfaces appears strongly dependent on the passivation of the initial Si surface, which drastically impacts on the spreading parameter.

## ACKNOWLEDGMENTS

This research was supported by the French National Research NUAGES Project (Grant no. ANR-21-CE24-0006). DFT calculations were performed at FOTON Institute, and the work was granted access to the HPC resources of TGCC/CINES under the allocation A0120911434 made by GENCI.